# Influence of driving frequency on the metastable atoms and electron energy distribution function in a capacitively coupled argon discharge

S. Sharma

*Institute for Plasma Research, Gandhinagar -382428, India*

*Homi Bhabha National Institute, Anushaktinagar, Mumbai 400094, India*

N. Sirse, M. M. Turner and A. R. Ellingboe

*Dublin City University, Dublin 9, Republic of Ireland*

One-dimensional particle-in-cell simulation is used to simulate the capacitively coupled argon plasma for a range of driving frequency from 13.56 MHz to 100 MHz. The argon chemistry set can, selectively, include two metastable levels enabling multi-step ionization and metastable pooling. The results show that the plasma density decreases when metastable atoms are included with higher discrepancy at higher excitation frequency. The contribution of multistep ionization to overall density increases with excitation frequency. The electron temperature increases with the inclusion of metastable atoms and decreases with excitation frequency. At lower excitation frequency, the density of $Ar^{**}$ ($3p^5$ 4p, 13.1 eV) is higher than $Ar^*$ ($3p^5$ 4s, 11.6 eV), whereas, at higher excitation frequencies the $Ar^*$ ($3p^5$ 4s, 11.6 eV) is the dominant metastable atom. The metastable and electron temperature profile evolve from a parabolic profile at lower excitation frequency to a saddle type profile at higher excitation frequency. With metastable, the *electron energy distribution function* (EEDF) changes its shape from Druyvesteyn type, at low excitation frequency, to bi-Maxwellian, at high frequency plasma excitation, however a three-temperature EEDF is observed without metastable atoms.

## I. INTRODUCTION

Capacitively coupled plasma (CCP) discharge driven at 13.56 MHz remains industrial standard tool for plasma processing applications including thin film depositions and reactive ion etching (RIE)[1]. A recent trend towards increase in the processing rates of CCP discharges is utilizing plasma excitation in very high frequency (VHF) range, i.e. from 30 – 300 MHz. Further benefits of VHF CCP discharges are low temperature processing, lower substrate damage and unique gas-phase chemistry[2-7]. Most recent example is the deposition of thin film of silicon nitride ($Si_3N_4$) on lithesome substrate for flexible organic electronic devices, where using 162 MHz plasma excitation a thin film of high optical transmittance and low WVTR is deposited at a substrate temperature of $100^0$ C[8-9]. In etching of S-MAP and organic low-$k$ film having an inorganic film etch mask, the 100 MHz radio-frequency (rf) RIE plasma process have shown great improvement of the carbon film etch profile, high selectivity, less erosion of the $SiO_2$ mask edge, and straight sidewall profile in comparison to 13.56 MHz plasma excitation[10].

Besides process improvements, the VHF plasma excitation showed significant difference in the discharge characteristics including electron heating mechanism when compared to 13.56 MHz CCP. For instance, the VHF excited CCP discharges produce high plasma density and low self-bias[11]. This is due to increase in the discharge current and plasma density at VHF plasma excitation for a constant discharge power. The VHF plasma excitations also exhibit a bi-Maxwellian type *electron energy distribution function* (EEDF), irrespective of gas pressure or gas type[12-15]. The transformation of EEDF to bi-Maxwellian at VHF was also predicted by particle-in-cell simulation[16]. The electron heating mechanism is mostly stochastic at a low gas pressure in 13.56 MHz CCP[17], whereas, it is observed that the CCP discharges driven at VHF are mostly produced and sustained by "beam like electrons" which are produced from near to the edge of expanding sheath[16, 18-19]. Furthermore, it is experimentally observed that the higher excitation frequency, 100 MHz in comparison to 60 MHz, yield higher densities of negative ion species at a constant rf power[20]. The enhancement in the negative ion densities with excitation frequency is further suited for etching of future generation micro-devices with a sub-nm characteristic size when using a pulsed plasma system[21].

Apart from neutrals and charged species, there exist long-lived metastable atoms and molecules in the discharge. These metastable are produced through various mechanisms but the important one is electron impact excitation of ground state. The presence of metastable causes obvious electron energy dissipation and affects the discharge characteristics significantly. However, the metastable existence could enhance plasma density through multi-step ionization and metastable pooling. In gas mixtures, the metastable collisions with other species further drive many other chemical transformations. In the past decades, many simulation, theoretical and experimental studies have been performed to investigate the influence of metastable on discharge characteristic in low to high pressure CCP discharge excited at 13.56 MHz[22-35]. However, there exist very few studies describing the effect of excitation frequency. Colgan et al[6] performed fluid simulation and experiment in capacitively coupled argon discharge in the frequency range of 13.56 MHz to 54.4 MHz and at a fixed gas pressure of 250 mTorr. This study was emphasized on the scaling law at a constant discharge voltage and discharge current. Hebner et al[35] measured optical emission and argon metastable density along with line-integrated electron density and ion saturation current in a dual-frequency, capacitively coupled, 300 mm-wafer plasma processing system excited in the frequency range from 10 MHz to 190 MHz. They discovered that the argon metastable density and spatial distribution were not a strong function of drive frequency. Zhang et al[36] used fluid simulation to simulate the capacitively coupled argon plasma including the metastable in the frequency range from 0.5 to 30 MHz at a gas pressure of 300 mTorr and 1 Torr. For both pressure conditions, and for all the excitation frequencies, they observed a saddle distribution of metastable profile in the axial direction.

The aim of this study to systematically investigate the influence of metastable atoms on the discharge properties including EEDF versus excitation frequency, 13.56 MHz – 100 MHz, in CCP discharges. We employed one-dimensional, particle-in-cell (PIC) code with Monte Carlo Collisions (MCCs). Although computationally expensive, the PIC, unlike fluid simulation, makes no implicit assumptions on the electron velocity distribution function and therefore provides useful insight of the plasma kinetics. The study is performed in a CCP with a discharge gap of 3.2 cm. The applied voltage and gas pressure is kept constant at 100 V and 100 mTorr respectively. We further examined the effect of excitation frequency on the axial distribution of metastable densities and EEDF.

## II. DESCRIPTION OF THE PIC SIMULATION

The simulation procedure, which is used in present research work is based on Particle- in-Cell/Monte Carlo collision (PIC/MCC) methods[37-38]. The main advantage of employing a PIC code is that there are no assumptions to the calculation of electron velocity distribution function (EEDF) implicitly, although the price to be paid is that a PIC code is computationally very expensive. We have used well-tested 1D3V, self- consistent, electrostatic, PIC code developed at Dublin City University, Ireland in the current research work[39]. This code was recently used to validate the analytical models predicting stochastic heating, field reversal and ion reflection from the sheath region, and electric field transients in 13.56 MHz CCP discharge[40-42]. The electron-neutral (elastic, inelastic, and ionization) and ion-neutral (elastic, inelastic, and charge exchange) collisions are considered for all set of simulations. Furthermore, charged particles, electrons and positive ions, and two lumped excited states of Ar i.e. Ar$^*$ (3p$^5$4s), 11.6 eV, and Ar$^{**}$ (3p$^5$4p), 13.1 eV, in uniform neutral argon gas background is considered. The cross-section data used here are from well-tested sources[43].

The detailed reactions considered here are listed in table 1.

**Table 1:** Collisions considered in the simulation of CCP argon discharges.

| Process | Reaction |
| --- | --- |
| 1. Elastic scattering | e + Ar --> Ar + e |
| 2. Ground state excitation to Ar$^*$ | e + Ar --> Ar$^*$ + e |
| 3. Ground state exciation to Ar$^{**}$ | e + Ar --> Ar$^{**}$ + e |
| 4. Ground state ionization | e + Ar --> Ar$^+$ + 2e |
| 5. Superelastic collisions of Ar$^*$ | e + Ar$^*$ --> Ar + e |
| 6. Superelastic collisions of Ar$^{**}$ | e + Ar$^{**}$ --> Ar + e |
| 7. Further excitation | e + Ar$^*$ --> Ar$^{**}$ + e |
| 8. Partial de-excitation | e + Ar$^{**}$ --> Ar$^*$ + e |
| 9. Multi-step ionization (Ar$^*$) | e + Ar$^*$ --> Ar$^+$ + 2e |
| 10. Multi-step ionization (Ar$^{**}$) | e + Ar$^{**}$ --> Ar$^+$ + 2e |
| 11. Charge exchange | Ar$^+$ + Ar --> Ar + Ar$^+$ |
| 12. Elastic scattering | Ar$^+$ + Ar --> Ar$^+$ + Ar |
| 13. Inelastic Ar$^*$ production | Ar$^+$ + Ar --> Ar$^*$ + Ar$^+$ |
| 14. Metastable pooling | Ar$^*$ + Ar$^*$ --> Ar$^+$ + Ar + e |
| 15. Ionization | Ar$^+$ + Ar --> Ar$^+$ + Ar$^+$ + e |
| 16. Elastic scattering | Ar$^*$ + Ar --> Ar$^*$ + Ar |
| 17. Elastic scattering | Ar$^{**}$ + Ar --> Ar$^{**}$ + Ar |

The simulation region, 3.2 cm electrode separation, is divided into 512 number of grids and number of particles per cell is 100 for all cases. The time step has been chosen of the order of $10^{-11}$ s. The electrodes are planar and parallel to each other with infinite dimension and secondary electron emission is also ignored. The electrodes are perfectly absorbing for both electrons and ions. One of the electrodes is grounded while an RF voltage having the following wave-form drives the other one:

$$V_{rf}(t) = V_0 \sin(2\pi f_{rf} t + \phi) \quad \text{---------------(1)}$$

The metastable or neutral excited states evolve on much slower time scales compared to electrons because the argon to electron mass ratio is greater than ~70,000, and therefore the simulation is run for ~5200 RF cycles to achieve the steady states for all cases[28]. We have considered three cases here: (1) Without metastable case, which includes equations 1, 4, 11, 12 and 15 from table 1 (2) With metastable including multistep ionization, which includes all set of equations in table 1. (3) With metastable no multistep ionization, which includes all set of equations except equation 9 and 10 in table 1.

**III. RESULTS AND DISCUSSIONS**

Fig. 1 shows the plasma density and electron temperature versus excitation frequency at the centre of the discharge for without metastable, with metastable-multistep ionization and with metastable-no multistep ionization. These parameters are deduced from the EEDF using, plasma density $\propto \int_0^{\varepsilon_{max}} F(\varepsilon)d\varepsilon$ and, $T_{eff} \propto \int_0^{\varepsilon_{max}} \varepsilon F(\varepsilon)d\varepsilon / \int_0^{\varepsilon_{max}} F(\varepsilon)d\varepsilon$, where $F(\varepsilon)$ is EEDF and $\varepsilon$ is the electron energy. As displayed in Fig. 1 (a), the plasma density increases with excitation frequency for all cases. However, there exist substantial difference in the absolute densities. At all excitation frequencies, the without-metastable case is producing the highest plasma density, whereas, with metastable-no multistep ionization the plasma density is the lowest. Including multistep ionization is boosting the plasma density when compared to no-multistep ionization but it is still lower than without metastable. Unlike plasma density, a reverse trend is observed for the electron temperature. As shown in Fig. 1 (b), the electron temperature decreases with increasing excitation frequency for with-metastable, whereas, it increases with frequency when the metastable are not considered in the simulation. The electron temperature is consistently lower for without metastable while it is highest for with metastable-no multistep ionization. Including multistep ionization, the electron temperature decreases slightly when compared to no multistep ionization, but its trend versus excitation frequency is same. These results are in contradict with previous fluid simulations[6, 36] which predict either constant or an increase in the electron temperature, however, our results are in good agreement with the experiment[12].

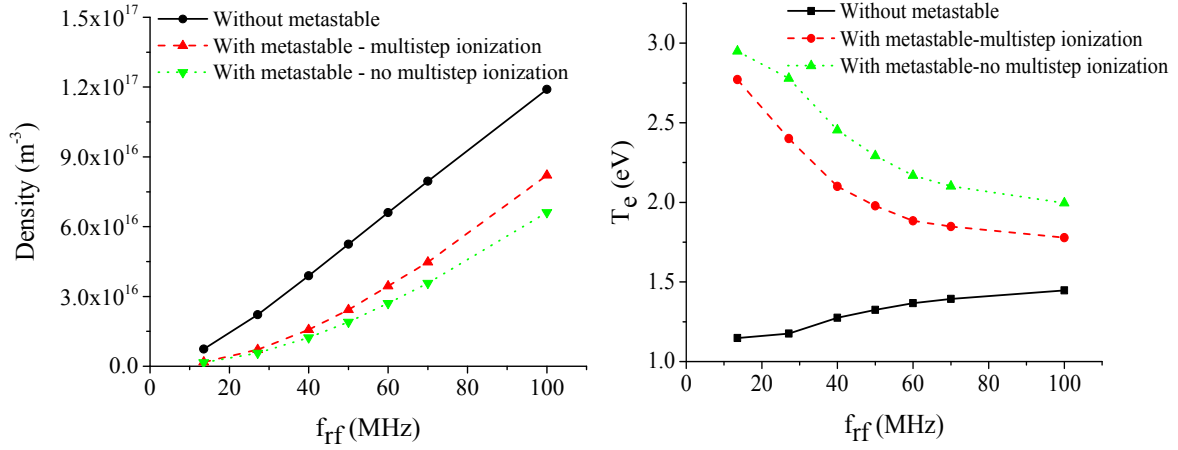

FIG. 1. (a) The plasma density and (b) electron temperature versus excitation frequency at the centre of the discharge for without metastable, with metastable-multistep ionization and with metastable-no multistep ionization.

An increase in the plasma density with excitation frequency is mainly attributed to increase in the discharge current. At constant discharge voltage, increase in the excitation frequency reduces sheath impedance which allows more current to flow through plasma, increasing the rf power into the plasma, and therefore plasma density increases. Without metastable case, we observed an increase in peak discharge current from 13 A/m$^2$ at 13.56 MHz to 360 A/m$^2$ at 100 MHz. Similarly with metastable – multistep ionization and no multistep ionization, the discharge current increases from 10 A/m$^2$ at 13.56 MHz to 300 A/m$^2$ and 280 A/m$^2$ at 100 MHz respectively. The increasing trend of plasma density versus excitation frequency is also in accordance with, $Plasma\ Density \propto \omega_{rf}^2$ as predicted for a constant voltage case[1]. A decrease in the plasma density with the inclusion of the metastable atoms is attributed to power losses in the several reactions including production and losses of metastable atoms, table 1. Furthermore at constant pressure, the production of metastable decreases the ground state atom density. Both of these decreases the direct electron impact ionization rates from the ground state and therefore the plasma density decreases. The inclusion of multistep ionization increases the ion-electron production channel which in turn increases the plasma density, however, it is still lower than without metastable case. It is noteworthy that the difference between plasma density for with metastable - multistep ionization and no multistep ionization increases with excitation frequency. At 13.56 MHz the percentage difference in plasma density is ~10 % which is increasing up to ~20 % at 100 MHz excitation frequency. These results show that the contribution of multistep ionization to plasma density is higher at higher excitation frequency. With metastable, an observed decreasing trend of electron temperature versus excitation frequency, Fig 1 (b), is attributed to an increase in the population of low energy electrons. This is mainly due to enhanced ionization at higher excitation frequency which produces low energy electrons. Figure 2 shows the EEDF at the centre of the discharge for different excitation frequencies from 13.56 – 100 MHz. As shown in Fig 2 (a), the low electron energy population increases with excitation frequency. The high energy population also increases, however, it increases at a lower rate than low energy electron population. Furthermore, a peak in the EEDF is observed to shift towards low energy. Due to change in the population of low and high energy electron population at a different rate transforming the shape of EEDF from Druyvesteyn to bi-Maxwellian which is described previously[16]. The transformation of EEDF from Druyvesteyn at low frequency to bi-Maxwellian at very high frequency was also predicted and in agreement with the experiments[12-15]. Without metastable, the density of both low energy electrons and tail electrons increases, albeit, nearly at the same rate (Fig 2 (b)), and therefore the electron temperature also increases.

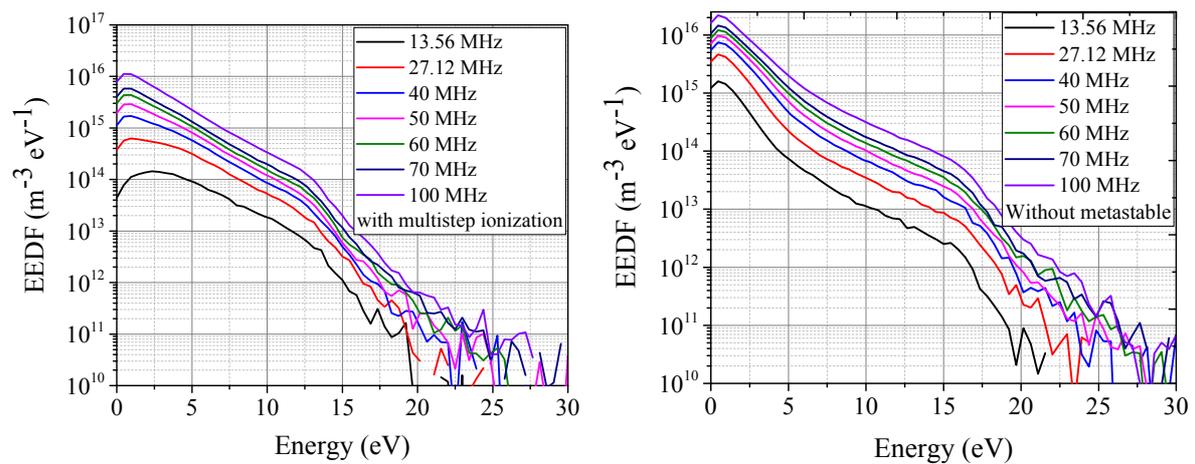

FIG. 2. EEDF at the centre of the discharge for different excitation frequencies from 13.56 – 100 MHz, with metastable – multistep ionization and without metastable atoms.

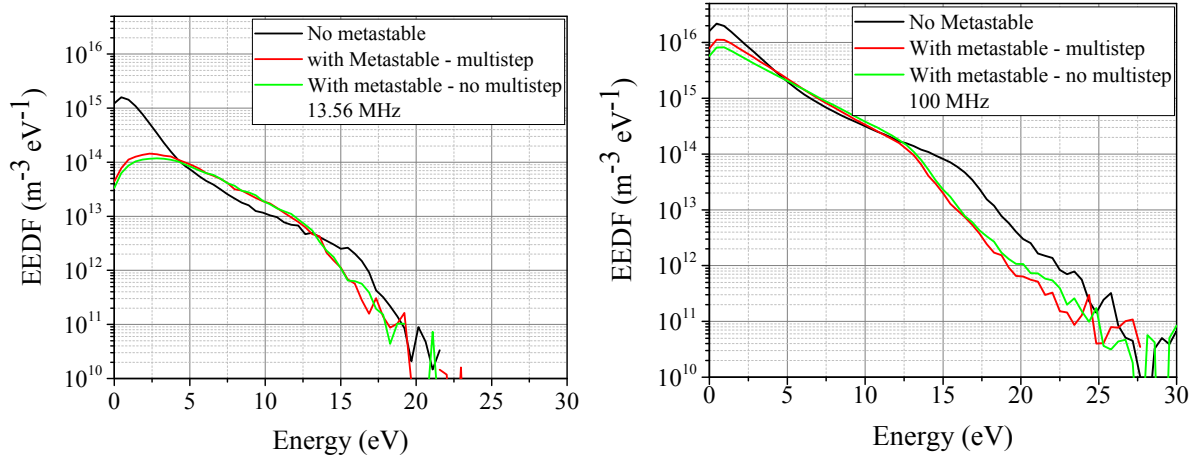

FIG. 3. EEDF at the centre of the discharge for without metastable, with metastable-multistep, with metastable-no multistep. (a) 13.56 MHz and (b) 100 MHz.

Figure 3 (a) and 3(b) plot the EEDF for each of the three cases, at excitation of 13.56 MHz and 100 MHz, respectively. As shown in Fig. 3 (a), when metastable are included the low energy electron population decreases and a depletion of the high-energy tail is observed. Similar results are obtained at 100 MHz excitation, Fig. 3 (b). A decrease in the population of low energy electrons and depletion of high-energy electrons in with-metastable case is mainly attributed to energy losses in the inelastic collisions. These processes significantly affect the ionization rate and therefore plasma density also decreases. It is interesting that the population of mid energy electrons, i.e. from ~ 4-13 eV, is higher in the case of with metastable when compared to without metastable. One of the reasons for observing higher population of these mid energy range electrons is the production of metastable atoms. When metastable are included, they are mostly formed by electron impact excitation of the ground state. In the case of Ar, the maximum collision cross sections of such reactions are observed at an electron energy of 20-30 eV[43], and therefore large number of mid energy range electrons are produced during metastable production process. An increase in the population of these mid energy electrons justifies higher electron temperature as observed in with metastable cases. As shown in Fig. 2 (b), the fraction of mid energy range electron is decreasing at 100 MHz and therefore the difference between electron temperature for with and without metastable decreases with excitation frequency.

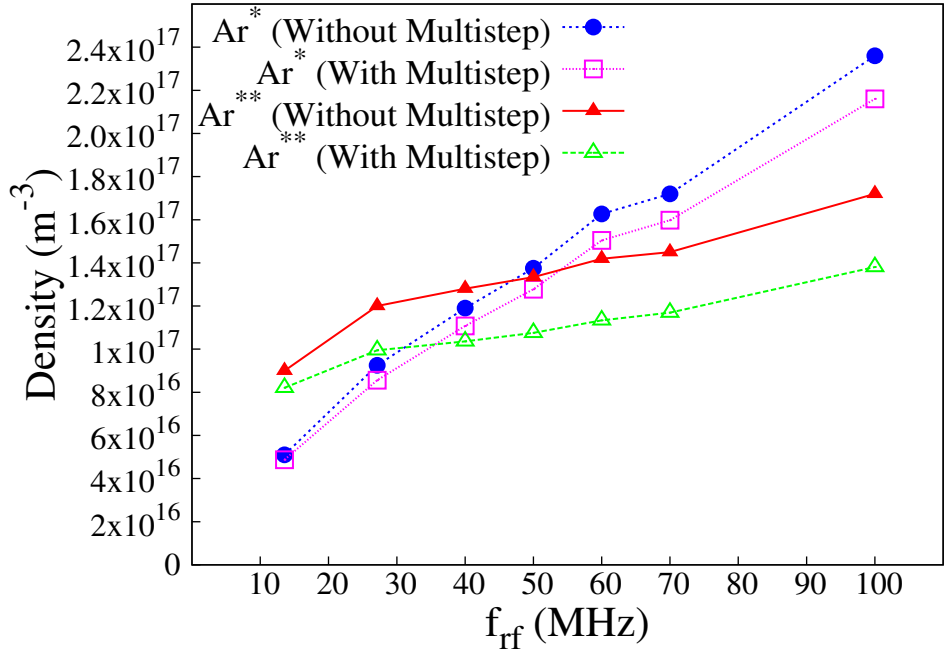

FIG. 4. The argon metastable densities versus excitation frequency at the centre of the discharge for with metastable-multistep ionization and with metastable-no multistep ionization.

Fig. 4 plots the densities of Ar$^*$ and Ar$^{**}$ at the centre of the discharge as a function of excitation frequency. The densities are shown for both with and without multistep ionization. As shown in the figure the densities of Ar$^*$ and Ar$^{**}$ increases with excitation frequency. For multistep ionization, it is observed that the density of Ar$^*$ increases from ~$5\times10^{16}$ m$^{-3}$ at 13.56 MHz to ~$2.2\times10^{17}$ m$^{-3}$ at 100 MHz, whereas, the density of Ar$^{**}$ increase from $9\times10^{16}$ m$^{-3}$ at 13.56 MHz to $1.4\times10^{17}$ m$^{-3}$ at 100 MHz. This observed increase in Ar$^*$ and Ar$^{**}$ densities is again due to increase in the discharge current with increase in the excitation frequency, which increases the discharge power. As expected, for without multistep ionization, the densities of Ar$^*$ and Ar$^{**}$ are consistently higher in comparison to with multistep ionization. It is important to note that there is a crossover in the densities of Ar$^*$ and Ar$^{**}$ with excitation frequency. At 13.56 MHz, Ar$^{**}$ is the dominant metastable atom, whereas, at 100 MHz the density of Ar$^*$ is higher in comparison to Ar$^{**}$. One of the possible reasons for this discrepancy is an overall decrease in the electron temperature with excitation frequency. The excitation energy of Ar$^*$ and Ar$^{**}$ are 11.6 eV and 13.1 eV respectively and therefore it very likely to produce low excitation energy Ar$^*$ metastable at 100 MHz due to lower electron temperature. The observed trend of metastable densities versus excitation frequency at the centre of the discharge may not be same at different locations in the discharge, and therefore we further examined the profile of metastable atom densities between the electrodes.

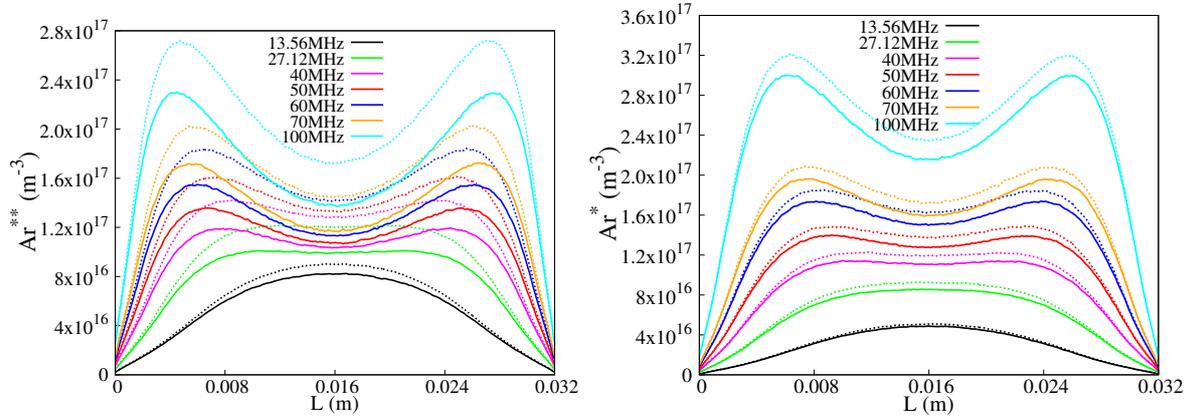

FIG. 5. The argon metastable profile a) Ar$^{**}$ and b) Ar$^*$ for different excitation frequencies. Dotted line and continuous lines are showing without and with multistep ionization respectively.

Fig. 5 shows the profile of Ar$^*$ and Ar$^{**}$ densities between the electrodes for different excitation frequencies. We observed that the metastable densities increase with excitation frequency at all locations between the electrodes. Without multistep ionization, the metastable densities are always higher. The Ar$^{**}$ density is higher than Ar$^*$ density at lower excitation frequency, whereas, at higher excitation frequency Ar$^*$ is the dominant metastable atom. It is interesting that the metastable profile changing its shape from parabolic, at 13.56 MHz, to a saddle type distribution at higher excitation frequencies. The transformation of the metastable profile from parabolic to saddle type is also observed in the case without multistep ionization. This confirm that the dip in the metastable densities in the bulk plasma at higher excitation frequency is not due to step-wise ionization but it is directly related to the excitation frequency which produces higher densities of metastable atoms. The saddle profiles were also observed by previous authors at 13.56 MHz excitation frequency and at higher gas pressures[26-28]. They attributed this effect to reduction in the mean free path of electrons at higher gas pressures resulted to enhanced excitation to metastable states near to the sheath edge. Since the production of metastable atoms is directly related with the electron temperature, therefore, we further examined the electron temperature profile to understand the observed saddle type profile at higher excitation. Fig. 6 shows the electron temperature profile for different excitation frequencies and with multistep ionization. As shown in the figure, a similar type of transition is observed in the electron temperature profile. At 13.56 MHz, the electron temperature is maximum at the centre of the discharge and decreasing towards the electrodes. At higher excitation frequency, off-centred peak in the electron temperature profile is observed. The variation of electron temperature between the electrodes represents a variation in the mean electron energy which is localized near to the sheath at higher excitation frequency. The source term of metastable states is a function of gas density, electron density and excitation rate coefficients which is a function of electron temperature. At constant gas pressure the neutral gas density is constant. Furthermore, as shown in Fig. 6 (b), the electron density profile is also parabolic at higher excitation frequency, and therefore the observed saddle profile of metastable densities is mainly attributed to non-uniform electron temperature profile.

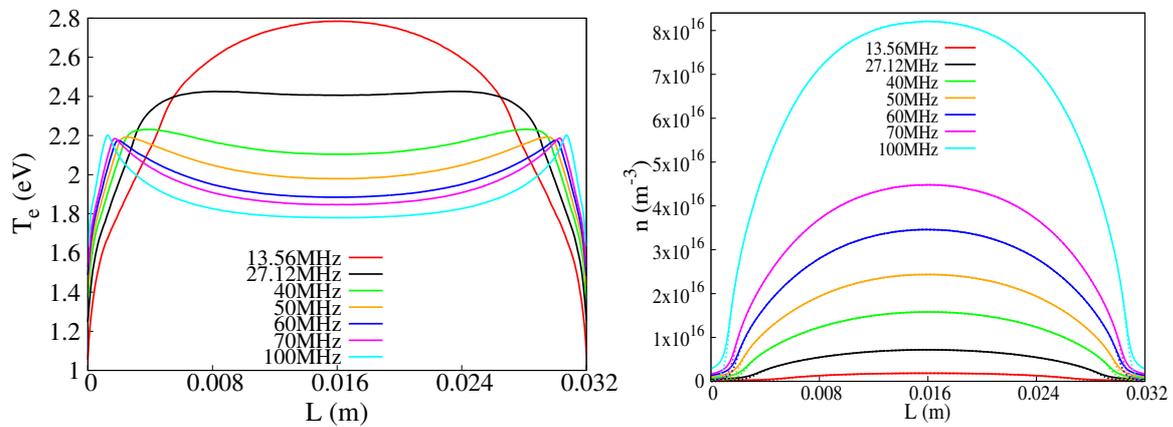

FIG. 6. The electron temperature between electrodes for different driving frequencies.

## IV. Summary and conclusions

We have performed one-dimensional, self-consistent, particle-in-cell simulation to investigate the effect of excitation frequency on the metastable atoms and electron energy distribution function in a low pressure capacitively coupled Ar plasma. The excitation frequency is varied from 13.56 MHz to 100 MHz, and the simulation is performed for 3 cases namely; without metastable, with metastable-no multistep ionization and with metastable-multistep ionization for a constant discharge voltage and at a fixed gas pressure of 100 mTorr.

It is shown that the plasma density decreases when metastable atoms are included. The difference is higher at higher excitation frequency. Including multistep ionization increases the plasma density when compared to without multistep ionization, however, it is still lower than without metastable case. These results suggest that the multistep ionization and metastable pooling doesn't play significant role in overall plasma density in comparison to ground state ionization. Meanwhile, the electron temperature is showing the reverse trend i.e. it is lower in the case of without metastable and higher in the case of with metastable-no multi step ionization. The EEDF comparison for 3 different cases shows that this discrepancy is due to the higher population of mid energy range electrons, 4-13 eV, which is possible due to the production of metastable atoms. Moreover, a change in the shape of EEDF is observed from Druyvesteyn type at lower excitation frequency to bi-Maxwellian at higher exciation frequency. An increase in the densities of both $Ar^*$ and $Ar^{**}$ is observed versus excitation frequency. At lower excitation frequency, the density of $Ar^{**}$ is higher than $Ar^*$, whereas, at higher excitation frequency $Ar^*$ is the dominant metastable atom. Additionally, the profile of metastable atoms is changing from parabolic shape at lower excitation frequency to a saddle type distribution at higher excitation frequency. This transformation is due to the change in the profile of electron temperature which is also showing off-centred peak at higher excitation frequency. The off-centred peak in the electron temperature profile is suggesting a heating mode transition which is subject of future publication.